\documentclass{emulateapj}
\usepackage{graphicx,amssymb}
\usepackage{subfigure,appendix,enumerate}
\usepackage{hyperref,url,xcolor}

\long\def\symbolfootnote[#1]#2{\begingroup%
\def\thefootnote{\fnsymbol{footnote}}\footnote[#1]{#2}\endgroup}

\definecolor{dark-green}{rgb}{0.1,0.49,0.4}
\hypersetup{colorlinks=true, urlcolor=blue, citecolor=dark-green}

\newcommand\qso{Q 2237+305}
\newcommand\rxj{RX J1131-1231}

\begin{document}

\shorttitle{Einstein Cross SMBH Spin}
\shortauthors{Reynolds et al.}

\title {A Rapidly Spinning Black Hole Powers the Einstein Cross}
 \author{Mark T. Reynolds\altaffilmark{1}, Dominic J. Walton\altaffilmark{2}, Jon
   M. Miller\altaffilmark{1}, Rubens C. Reis\altaffilmark{1}}
                                                             
\email{markrey@umich.edu}

\altaffiltext{1}{Department of Astronomy, University of Michigan, 311 West Hall, 1085
  S. University Ave, Ann Arbor, MI 48109}
\altaffiltext{2}{Cahill Center for Astronomy and Astrophysics,
  California Institute of Technology, Pasadena, California 91125, USA}


\begin{abstract}
Observations over the past 20 years have revealed a strong relationship between the
properties of the supermassive black hole (SMBH) lying at the center of a galaxy and the
host galaxy itself. The magnitude of the spin of the black hole will play a key role in
determining the nature of this relationship. To date, direct estimates of black hole spin
have been restricted to the local Universe. Herein, we present the results of an analysis
of $\sim$ 0.5 Ms of archival \textit{Chandra} observations of the gravitationally lensed
quasar \qso~(aka the ``Einstein-cross''), lying at a redshift of z = 1.695. The boost in
flux provided by the gravitational lens allows constraints to be placed on the spin of a
black hole at such high redshift for the first time. Utilizing state of the art
relativistic disk reflection models, the black hole is found to have a spin of $a_* =
0.74^{+0.06}_{-0.03}$ at the 90\% confidence level. Placing a lower limit on the spin, we
find $a_* \geq 0.65$ (4$\sigma$).  The high value of the spin for the $\rm \sim
10^9~M_{\sun}$ black hole in \qso~lends further support to the coherent accretion
scenario for black hole growth. This is the most distant black hole for which the spin has
been directly constrained to date.
\end{abstract}
 

\maketitle
\section{Introduction}
Black holes play a key role in the growth and evolution of galaxies, and their stellar
content \citep{fabian12}.  The spin of the black hole ($a_* \equiv \rm Jc/GM^2$) is
crucial as it can influence both the radiative and kinetic components of the energy
output. In an optically thick geometrically thin accretion disk the radiative
efficiency, and hence luminosity ($L=\eta\dot{M}c^2$), of the inner region depends on the
spin, peaking at $\eta \gtrsim 0.3$ for a maximal spin black hole ($a_* \sim 0.998$,
\citealt{thorne74}).  This should be compared to the fiducial spin value estimated from
observations of the cosmic X-ray background (CXB, $\eta \sim 0.1$ or equivalently $a_*
\sim 0.67$, \citealt{soltan82}). Similarly, if relativistic jets are powered by tapping
the spin of the black hole via, for example the ``Blandford-Znajek'' process, a larger
spin may result in a more powerful jet ($\rm P_{jet} \propto$ $a_*^2\Phi^2$, where $\Phi$
is the magnetic flux, \citealt{bz77}). Hence, the spin distribution of black holes is a
crucial ingredient in the feedback process effecting the co-evolution of the black hole
and its host galaxy \citep{fabian12}.

Direct constraints on the spin of a black hole are now possible via modeling the accretion
disk reflection spectrum \citep{lightman88,fabian89,tanaka95,miller07}. Detailed
observations of AGN in the local universe ($\rm z \lesssim 0.1$) illustrate the power of
this method to constrain the inner accretion flow geometry in the strong GR regime through
both spectral (e.g., \citealt{fabian09,risaliti13,reynolds13}) and timing methods (e.g.,
\citealt{zoghbi12,demarco13,kara13,uttley14}). However, at the current time it is only
possible to indirectly probe the spin distribution for black holes at high redshift
\citep{davis11,wu13,trakhtenbrot14}, though such methods are necessarily hampered by
substantial systematic uncertainties, e.g., \citet{raimondo12}.

In order to determine the influence of the black hole on structure growth we would like to
probe the spin distribution out to the epoch of peak galaxy and star formation ($\rm z
\sim 2$). In \citet{reis14}, we demonstrated how the boost in flux provided by a strong
gravitational lens can be used to constrain the spin of a SMBH lying at a cosmologically
relevant distance, i.e., \rxj~at z = 0.658 ($\rm \tau_{lookback} \sim 6~Gyr$). The black
hole was found to have a large spin, $a_* = 0.87^{+0.08}_{-0.15}$ at the 3$\sigma$
confidence level. This is the most distant SMBH for which the spin of the black hole has
been directly measured to date. The high spin value would support the coherent accretion
scenario (e.g., \citealt{volonteri13}). This source class opens up a promising avenue to
begin to constrain SMBH spin evolution as a function of redshift, and will facilitate
comparison with black hole host-galaxy co-evolution models, in particular as models become
increasingly sophisticated, e.g., \citet{dubois14,sesana14}.

\begin{figure}
\begin{center}
\includegraphics[width=0.46\textwidth]{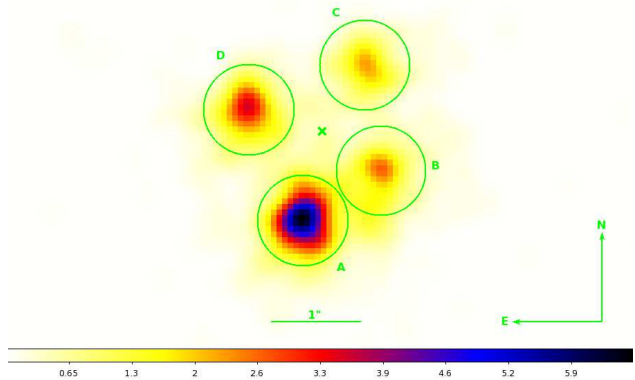}
\caption{Sample \textit{Chandra} image of \qso~(obsid: 14514) in the 0.35 -- 7.0 keV band,
  where the 4 lensed images are clearly resolved. The raw image has been rebinned to
  1/8$\rm^{th}$ the native pixel size before smoothing with a Gaussian ($\sigma =
  0.25\arcsec$). The exposure time is 29.36 ks. The green circles denote our source
  extraction regions ($\rm r=0.5\arcsec$), while the position of the lensing galaxy ($\rm
  z = 0.0395$) is marked by the cross.}
\label{cxo_image}
\end{center}
\end{figure}

\qso~was discovered by \citet{huchra85} and determined to be a strongly lensed quasar
lying at a redshift of z = 1.695. The quasar is lensed by a Sab barred galaxy at z =
0.0395 into 4 distinct components \citep{yee88}. Due to the low redshift of the lens,
\qso~commanded immediate interest due to the short inter image time-delays
($\Delta\tau_{ac}$ is $\sim$ 3 hr) and the discovery of micro-lensing, which facilitate
detailed study of the quasar and lensing galaxy (see Fig. \ref{cxo_image};
\citealt{irwin89,wambsganss94}). Optical spectroscopy of the gravitationally distorted
quasar host galaxy has resulted in the detection of broad and narrow line region
components consistent with emission from a typical Seyfert galaxy, e.g.,
\citet{motta04}. The black hole has been constrained to have a mass of $\rm log_{10}M_{BH}
= 9.08\pm0.39~M_{\sun}$, based on analysis of the H$\beta$ emission line
\citep{assef11}. In addition to being quadruply lensed, the quasar flux is also magnified
with an average magnification $\sim$ 16 \citep{schmidt98}.

The system is of significant interest at X-ray energies as micro-lensing can provide
constraints on the inner accretion flow at AU scales \citep{chartas12}. The quasar is
viewed almost face-on \citep{poindexter10,sluse11}. First resolved at X-ray wavelengths by
\textit{Chandra}, where in addition to the detection of an iron K line, a time delay at
X-ray energies was also detected ($\rm \Delta\tau_{ac} \sim 2.7~hr$; \citealt{dai03}). The
micro-lensing studies have also enabled constraints on the size of the corona, where the
source of the hard X-ray emission has been constrained to be compact with a size $\rm \sim
20~R_g$ or $\rm \sim 2.9 \times 10^{15}~cm$ \citep{mosquera13}, and the temperature
profile of the accretion disk, which is found to be consistent with expectations from
standard thin accretion disk theory, i.e., $\rm T(r) \propto r^{-3/4}$,
\citealt{morgan10}).

In this paper, we describe analysis of archival \textit{Chandra} observations of \qso,
where we directly measure the spin of the SMBH lying at a redshift of $\rm z \sim 1.7$.

\section{Observations}
Our sample consists of 26 observation in total, obtained over the first 13 yrs of
operation of \textit{Chandra}, i.e., ut000906 -- ut130106 where the relevant observation
IDs are: 00431, 01632, 06831--06840, 11534--11539, 12831, 12832, 13191, 13195, 13960,
13961, 14513, 14514. Analysis of the majority of these observations have been previously
presented in the context of micro-lensing studies of \qso~by \citet{dai03,chen12}
and collaborators. With the exception of the first 2 observations, which were taken with
the standard full frametime of 3.24s, all subsequent observations were obtained with a
frametime of 1.74s.  The summed exposure time is 468 ks.

\begin{figure}
\begin{center}
\includegraphics[width=0.46\textwidth]{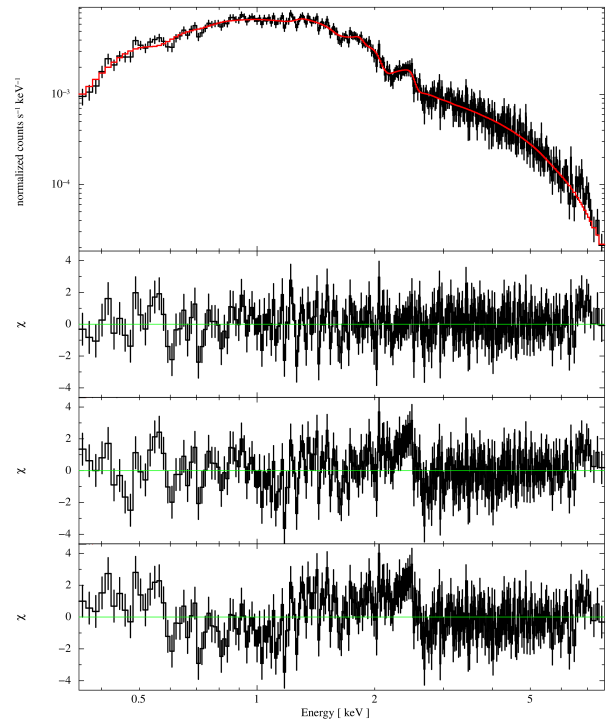}
\caption{Our initial phenomenological fit to the combined data. The final best fit model
  is a combination of a power-law, a disk component representing the `soft-excess' and a
  broad relativistically skewed iron line, i.e. \texttt{pha(zpha(diskbb+po+relline))}. The
  residuals to the best fit model are displayed below the upper panel and represent from
  bottom to top: a power-law alone, a power-law plus disk, and the final model of a disk
  plus line plus power-law. The best fit spin returned by this model is $a_* =
  0.75\pm0.03$ at the 90\% confidence level.}
\label{phenom_model}
\end{center}
\end{figure}

\begin{figure*}
\begin{center}
\subfigure{\includegraphics[width=0.34\textwidth,angle=-90]{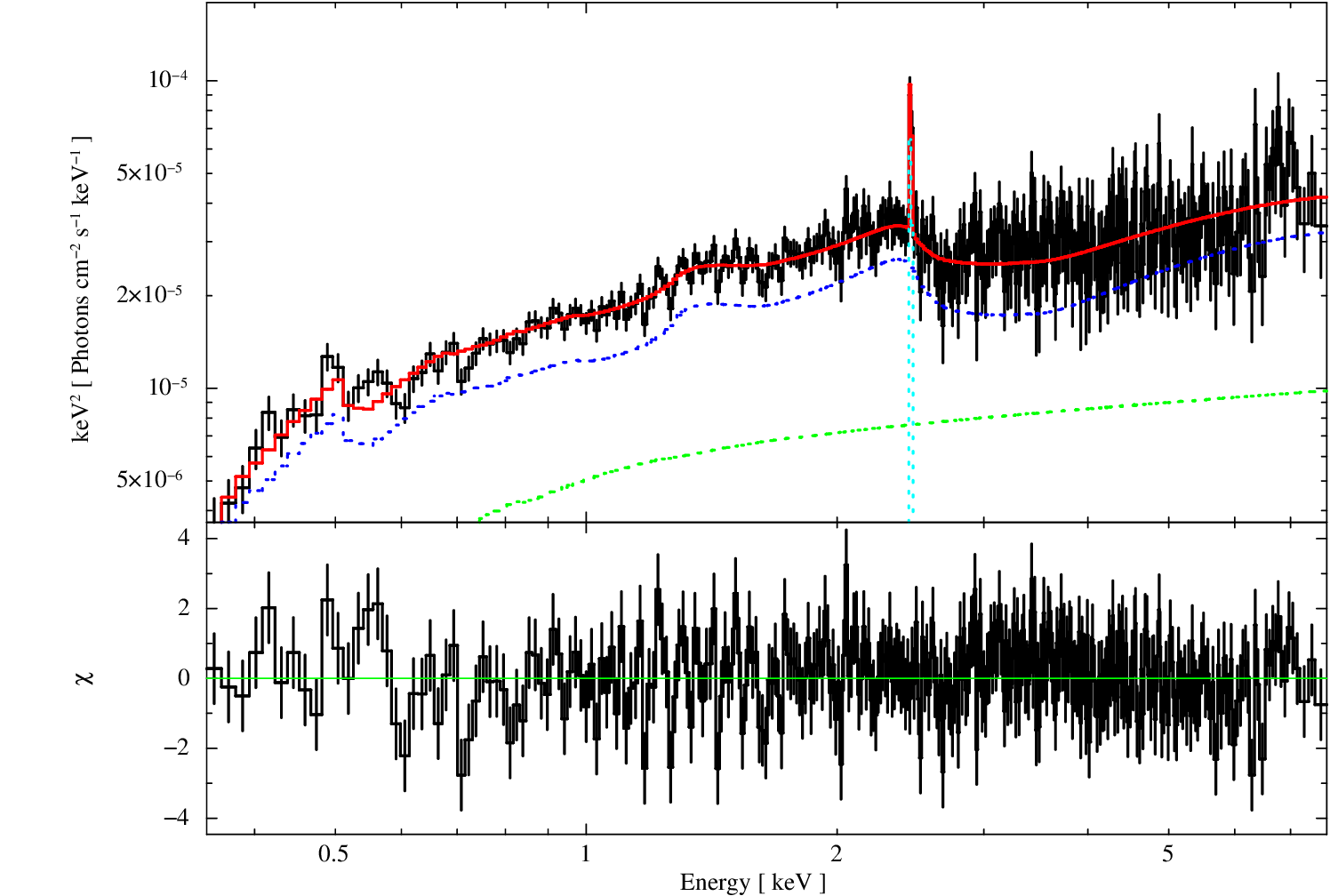}}
\subfigure{\includegraphics[width=0.34\textwidth,angle=-90]{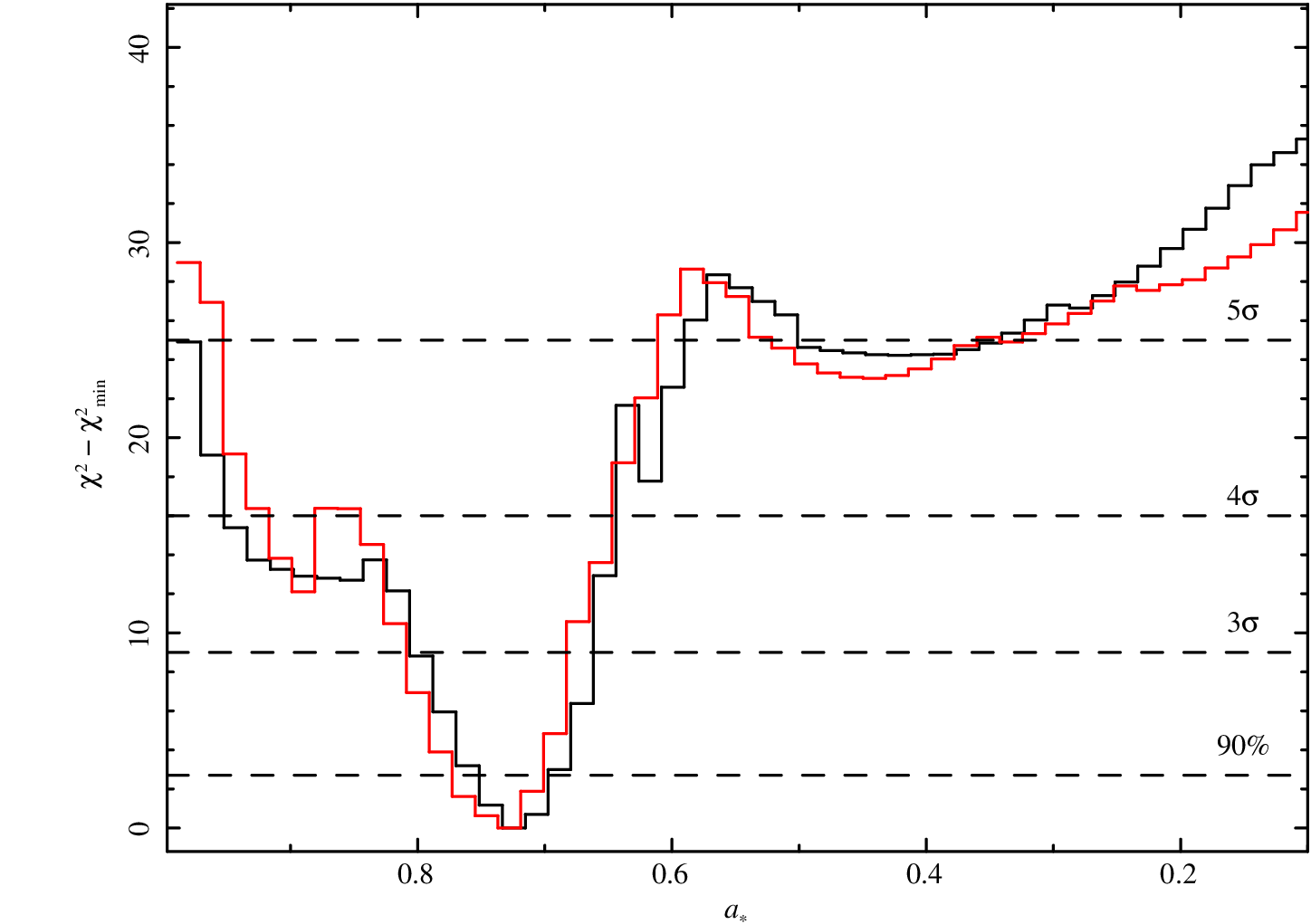}}
\caption{\textbf{Left:} The final best fit solar abundance self consistent relativistic
  reflection model (\texttt{pha(zpha(zpo+zgauss+relconv$*$reflionx))}) to the summed
  \textit{Chandra} spectrum of \qso~with residuals. The illuminating power-law is
  indicated in green, the reflected emission in blue and the narrow Fe K line in cyan,
  with the sum indicated in red. The black hole is determined to be rapidly spinning,
  i.e., $a_* = 0.73^{+0.05}_{-0.02}$ at the 90\% confidence level. \textbf{Right:} Contour
  plot of the spin parameter ($a_*$) calculated via the \texttt{steppar} command, for the
  reflection model plotted on the left (black: $\chi^2_{min} = 388$) and for a similar
  model with a variable iron abundance (red: $\chi^2_{min} = 371$). The 90\%, 3$\sigma$,
  4$\sigma$ and 5$\sigma$ confidence intervals are indicated by the dashed lines,
  demonstrating the robust nature of the spin constraint, i.e., $a_* \geq
  0.65~(4\sigma)$. See Table \ref{model_table} for the best fit parameters.}
\label{spectral_fits}
\end{center}
\end{figure*}

The analysis procedure follows that undertaken in our analysis of the lensed quasar
\rxj~\citep{reis14}.  In brief, all data are re-processed in \textsc{ciao
  v4.5}\footnote{\url{http://cxc.harvard.edu/ciao}} utilizing the latest \textit{Chandra}
calibration files in the standard manner. Although the individual components of \qso~are
separated by mere arc seconds, the unique spatial resolution of \textit{Chandra}
facilitates straightforward imaging of this lensed quasar at X-ray energies. Here
advantage is taken of the native sub-arcsecond spatial resolution provided by the
\textit{Chandra} mirrors and now available via the \textsc{EDSER} algorithm in
\textsc{ciao}. All images are re-binned to 1/8$\rm^{th}$ of the native ACIS pixel size before
smoothing with a Gaussian of 0.25$\arcsec$. An example image is displayed in
Fig. \ref{cxo_image}. Spectra are extracted from 0.5$\arcsec$ radius regions via the
\textsc{specextract} script with the PSF correction enabled, where the regions are
re-centered on each individual component in each image prior to extraction. The overall
flux from each individual lensed image from each observation was examined to determine if
any of the observations are affected by
pile-up\footnote{\url{http://cxc.harvard.edu/ciao/download/doc/pileup_abc.pdf}}.
Unsurprisingly given the cosmological distance, pile-up does not affect any of the
observations considered herein.  For completeness, the solitary \textit{XMM-Newton}
exposure obtained in 2002 is also considered \citep{fedorova08}.

All spectra are subsequently combined using the \textsc{combine\_spectra} script before
grouping to a signal-to-noise ratio of 3 per spectral bin with \textsc{dmgroup}. The
spectra and background files are then exported to \textsc{xspec}
v12.8.0m\footnote{\url{http://heasarc.gsfc.nasa.gov/xanadu/xspec/}} for spectral analysis.

\section{Analysis \& Results}
The final summed \textit{Chandra} spectrum contains data from 26 observations with 4
lensed images of the source per observation resulting in a total on source time of $\sim$
1.8 Ms. The resulting spectrum contains approximately 22 k net counts in the 0.35 -- 8.0
keV energy range. Data outside this interval are not considered further due to low S/N
and/or high backgrounds.

Initially, the brightest source image in each observation was characterized with a simple
power-law in order to search for evidence of extreme micro-lensing variability, which can
affect the Fe K region by, for example, introducing narrow red-shifted Fe line components
to the spectrum, such as that previously observed in \rxj~\citep{chartas12}. The brightest
image does not reveal any evidence for any anomalous variability; although, we note that
the S/N is necessarily low in individual exposures.

Our energy range of interest 0.35 -- 8 keV corresponds to the energy range 0.94 -- 21.56
keV in the quasar rest frame, thus providing access to the Fe K line and a portion of the
Compton hump region of the reflection spectrum. To characterize the spectrum, we fit the
data with a simple power-law modified by absorption in the Galaxy and the lens, i.e.,
\texttt{pha(zpha(zpo))}\footnote{The abundance model of \citet{asplund09} is used
  throughout this work, i.e., \texttt{aspl}.} where $\rm N_{H, local} = 0.05 \times
10^{22}~cm^{-2}$ and $\rm N_{H, lens} \sim 0.1 \times 10^{22}~cm^{-2}$ \citep{dai03}.
This fit reveals characteristic residuals consistent with those expected from the soft
excess at lower energies ($\lesssim$ 1 keV) and a broad iron line at a rest-frame energy
of $\sim$ 6.4 keV, see Fig. \ref{phenom_model}. In the standard manner, we initially
characterize the properties of the observed residuals using simple models, e.g., see
\citet{reis14}.

The soft excess is accounted for using a \texttt{diskbb} component. Fitting the broad line
component with a Gaussian reveals a broad line (EW = 167$\rm ^{+48}_{-42}~eV$, $\rm \sigma
= 0.52\pm0.14~keV$) whose centroid energy is skewed to lower energies ($\rm E_{line} =
6.05^{+0.20}_{-0.17}~keV$), as it attempts to account for the red wing of what is in fact
a relativistically broadened line. To more accurately model this line, we utilize the
relativistic line model \texttt{relline} of \citet{dauser10}. The final phenomenological
model (\texttt{pha(zpha(diskbb+po+relline))}) provides an excellent fit to the data
($\chi^2/\nu = 368/339$) as is evidenced in Fig. \ref{phenom_model}. Of particular note,
the \texttt{relline} component suggests a rapidly spinning black hole viewed at low
inclination, i.e., $a_* = 0.76\pm0.03$, $\rm E_{line} = 6.82^{+0.12}_{-0.09}~keV$, $\rm
q_{in} \geq 6.63$, $\rm r_{break} = 4.36^{+0.45}_{-0.31}~R_g$, $\rm q_{out} =
3.12^{+0.20}_{-0.26}$, and $\rm i \leq 13.08\degr$ where all of the errors are at the 90\%
confidence level. The low inclination returned by the model agrees with previous work
suggesting the quasar is viewed approximately face-on \citep{poindexter10,sluse11}. The
radial emissivity profile of the reflected emission is parameterized by a broken
power-law, where the large inner index is consistent with preferentially beamed emission
from a compact corona, lying at small separation from the black hole, onto the inner disk
due to a rapidly spinning black hole, e.g., see \citet{wilkins12}.

Given the clear presence of reflection signatures in the spectrum, we now proceed to apply
a self consistent disk reflection model in combination with a state of the art
relativistic convolution kernel: \texttt{relconv} and \texttt{reflionx}
\citep{dauser10,rossfabian05}. Our final model is \texttt{pha(zpha(zpo + zgauss +
  relconv*reflionx))}\footnote{Our results using the \texttt{reflionx} reflection tables
  were compared to similar fits using the recent \texttt{xillver} reflection tables of
  \citet{garcia13}, and were found to be consistent within the quoted errors.}. As this
model is physically self-consistent, an additional narrow Gaussian line is required to fit
the observed line profile ($\rm E_{line} \sim 6.6~keV$, EW $\sim$ 40 eV), see
Fig. \ref{spectral_fits}. This differs from the simple model above where the
\texttt{relline} component is free to fit the data without knowledge of the broader
continuum. Nonetheless, the parameters of the inner accretion flow are remarkably similar
to the phenomenological model, i.e., $a_* = 0.73^{+0.05}_{-0.02}$ (90\% confidence
level). This model is reflection dominated, $\rm R_{frac} = 2.8^{+2.8}_{-1.0}$, where the
reflection fraction is defined $\rm R_{frac} \equiv F_{refl}/F_{illum}$, see Table
\ref{model_table} for details.
 
The solar abundance model is of slightly lower quality in comparison to the
phenomenological model presented earlier, with evidence of systematic residuals at low
energy (compare Fig. \ref{phenom_model} \& \ref{spectral_fits}).  Enhanced iron abundances
have been discovered in the inner accretion disk for a number of AGN, e.g.,
\citet{fabian09,risaliti13}. As such, the above model is re-fit with a variable iron
abundance ($\rm \Delta\chi^2/\Delta\nu = 17$). This model is qualitatively similar to the
solar abundance model, see Table \ref{model_table}. However, the enhanced reflection
component in the iron band results in a modified continuum requiring a mildly harder
illuminating power-law ($\Gamma \sim 1.6$) and commensurately higher ionization of the
accretion disk ($\rm \xi \sim 10^3~erg~cm~s^{-1}$). The iron abundance is large, $\rm
Z_{Fe} \geq 6.2~Z_{\sun}$ (90\%) reducing to $\rm \sim 3.8~Z_{\sun}$ at the 3$\sigma$
confidence level.  In Fig. \ref{spectral_fits} (right), we plot confidence contours of the
spin parameter for both models. It is clear that our ability to constrain the spin is
robust with respect to our ability to constrain the Fe abundance.

The 2 -- 10 keV rest frame luminosity is measured to be $\rm L_x \sim 1.3 \times
10^{45}~erg~s^{-1}$ (Table \ref{model_table}). Given that the bolometric correction and
the magnification of this quasar work in the opposite sense and likely have similar
magnitudes (within a factor of 2, i.e., \citealt{schmidt98,vasudevan07}), this luminosity
corresponds to a relatively low Eddington rate of $\rm \sim
0.01~L_{Edd}~(10^9~M_{\sun}/M)$.

Finally, we note that the above model was also applied to the single existing
\textit{XMM-Newton} observation \citep{fedorova08}. This observation resulted in only
$\sim$ 12 ks goodtime with EPIC/pn and $\sim$ 20 ks for each of the MOS CCDs. The S/N in
this observation is low, containing only $\sim$ 2000 counts in the EPIC/pn exposure, and
$\sim$ 1000 with each MOS camera. Nonetheless, the spectrum is consistent with the
reflection spectrum detected by \textit{Chandra} at higher S/N discussed above. As such,
we do not consider this dataset further.

\begin{table}
\begin{center}
\caption{Best fit model parameters}\label{model_table}
\begin{tabular}{lcc}
\tableline\\ [-2.0ex]
Parameter & $\rm Z_{Fe} \equiv Z_{\sun}$ & $\rm Z_{Fe}~variable$ \\ [0.5ex]
\tableline\tableline\\ [-2.0ex] 
$\rm N_{H, local}~[10^{22}~cm^{-2}~]$ & 0.05 & 0.05\\ [0.5ex]
$\rm N_{H, lens}~[10^{22}~cm^{-2}~]$  & (0.11 $\pm$ 0.01)  & (0.06 $\pm$ 0.01)\\ [0.5ex]
$\rm E_{line}~[~keV~]$    & 6.58 $\pm$ 0.03  & 6.58 $^{+0.02}_{-0.06}$\\ [0.5ex]
$\rm Norm_{line}$ & (6.6 $\pm$ 2.9)$\times 10^{-7}$ & (6.6 $\pm$ 3.1)$\times 10^{-7}$\\ [0.5ex]
$\rm \Gamma$     & 1.80$\rm ^{+0.16}_{-0.11}$ & 1.64$\pm$0.04\\ [0.5ex]
$\rm Norm_{po}$   & (3.9$\rm ^{+1.6}_{-1.8}$)$\times 10^{-5}$ & (7.8$\pm$0.08)$\times 10^{-5}$\\ [0.5ex]
$\rm q_{in}$  & $\geq 7.6$  & $\geq 7.3$\\ [0.5ex]
$\rm q_{out}$  & 3.02$\pm$0.25 & 3.31$\pm$0.28\\ [0.5ex]
$\rm r_{break}~[~R_g~]$    & 4.65$\rm ^{+0.48}_{-0.32}$  & 4.35$\rm ^{+0.59}_{-0.26}$\\ [0.5ex]
$\rm i~[~\degr~]$           & $\leq 11.5$  & $\leq 11.7$\\ [0.5ex]
$a_*$          & 0.73$\rm ^{+0.05}_{-0.02}$ & 0.74$\rm ^{+0.06}_{-0.03}$\\ [0.5ex]
$\rm \xi~[~erg~cm~s^{-1}~]$        & 681$\rm ^{+339}_{-139}$ & 993$\rm ^{+155}_{-378}$\\ [0.5ex]
$\rm Z_{Fe}~[~Z_{\sun}~]$           & 1.0  & $\geq 6.2$\\ [0.5ex]
$\rm Norm_{reflionx}$ & (8.5$\rm ^{+7.1}_{-1.4}$)$\times 10^{-9}$ & (2.8$^{+4.5}_{-1.1}$)$\times 10^{-9}$\\ [0.5ex]
--- & --- & ---\\ [0.5ex]
$\rm R_{frac}$  & 2.4$^{+2.8}_{-1.0}$  & 0.40$\pm$ 0.12\\ [0.5ex] 
$\rm L_{2-10~keV}~[~erg~s^{-1}~]$ & $(1.26^{+0.95}_{-0.14})\times10^{45}$ & $(1.27^{+0.13}_{-0.54})\times10^{45}$\\ [0.5ex]
--- & --- & ---\\ [0.5ex]
$\chi^2/\nu$   & 388/339 & 371/338 \\ [0.5ex]
\tableline
\end{tabular}
\tablecomments{Parameters of our best fit model:
  \texttt{pha(zpha(zpo+zgauss+relconv$*$reflionx))}. Galactic absorption and absorption in
  the low redshift lensing galaxy are included. The reflection fraction is defined as,
  $\rm R_{frac} \equiv F_{refl}/F_{illum}$, where the unabsorbed fluxes are calculated in
  the 0.1 -- 100 keV band (observed frame) by extrapolation of the best fit model. The 2
  -- 10 keV rest frame luminosity is calculated via the \texttt{lumin} command
  assuming z=1.695, $\rm H_0 = 70~km~s^{-1}~Mpc^{-1},~\Omega_\Lambda = 0.73$. All
  parameters are quoted with errors at the 90\% confidence level.}
\end{center}
\end{table}

\section{Discussion}
In this paper we present a direct measurement of the spin of the black hole at high
redshift, i.e., $\rm \tau_{lookback} \sim 10~Gyr$. This is the most distant black hole for
which a direct constraint on the spin has been made to date. The spin of the black hole in
the quasar \qso~is found to be $a_* \geq$ 0.65 at the 4$\sigma$ confidence level. Thus the
\qso~SMBH is rapidly rotating during the epoch of peak star formation and galaxy
growth. Models for the spin evolution of a growing SMBH predict the largest differences
for the most massive black holes ($\rm M_{BH} \gtrsim 10^9~M_{\sun}$, e.g.,
\citealt{volonteri13}).  As a probe of SMBH/galaxy co-evolution, the measured high spin
for the black hole in \qso~would support the coherent accretion scenario for black hole
growth. It is also noteworthy that the high measured value for the spin implies an
accretion efficiency consistent with or greater than that implied by the Soltan argument
\citep{soltan82}.

Our analysis is necessarily aggressive given the moderate number of counts
available. However, the theoretical and observational support for the reflection model
utilized herein has considerable support from an increasing number of high S/N ratio
observations at low redshift, e.g.,
\citet{fabian09,zoghbi12,demarco13,risaliti13,kara13,walton14}.  The best fit reflection
model requires an enhanced iron abundance ($\rm Z_{Fe} \geq 3.8~Z_{\sun}~(3\sigma)$, see
Table \ref{model_table}). Such a large iron abundance in the inner accretion flow has been
observed in a number of local Universe AGN, e.g., 1H0707-495 \citep{fabian09}, NGC 1365
\citep{risaliti13}, and in larger samples of systems \citep{walton13}.

The enhanced iron abundance is required by the presence of residuals at low energies,
i.e., $\lesssim$ 1 keV in the observer frame. This corresponds to rest frame energies
$\lesssim$ 2.7 keV. Such residuals can in principle originate in an absorption component
\citep{miller08}, though a large column would be required to effect emission at such high
energies. This possibility was investigated by the addition of a partial covering absorber
to the solar abundance model, i.e.,
\texttt{pha(zpha(zpcfabs(zpo+zgauss+relconv*reflionx)))}. The improvement in the fit is
similar to the variable abundance model but with an extra degree of freedom (($\rm
\Delta\chi^2/\Delta\nu = 15/2$)). The column density is large as expected, $\rm N_H \sim
10^{23}~cm^{-2}$, and a covering fraction $\rm f_{cov} \sim 0.3$. This model requires a
soft power-law index of $\Gamma \sim 2$. This model is disfavored for a number of reasons:
(i) optical studies of \qso~find the extinction to be modest ($\rm 0.8 \lesssim A_V
\lesssim 1.2$, \citealt{agol00}) and consistent with the small X-ray column i.e., $\rm N_H
~\sim 10^{21}~cm^{-2}$ \citep{dai03}, (ii) optically thick winds are unlikely to form in
the inner disks of AGN at the luminosity observed herein ($\rm L_x \lesssim 0.01~L_{Edd}$,
\citealt{reynolds12}), and (iii) the recent \textit{NuStar} observations of low redshift
AGN, which have demonstrated the broadband spectrum to be consistent with the reflection
scenario, e.g., \citet{risaliti13,walton14}.

\qso~is the second high redshift quasar that we have been able to measure the spin of by
taking advantage of the boost in flux provided by a strong gravitational lens. Previously,
the spin of the $\rm log_{10}M_{BH} \sim 8.3~M_{\sun}$ SMBH in the z = 0.658 quasar
\rxj~has been determined to be $a_* = 0.87^{+0.08}_{-0.15}$ \citep{reis14}, also
consistent with the coherent accretion scenario.

Models for black hole growth and evolution and their co-evolution with the host galaxy are
becoming more sophisticated, e.g., \citet{dubois14,sesana14}. However, it is important to
note that caution is required in interpreting these observational results at the current
time. Apart from the small size of the sample, the quasars for which this type of study
will be possible do not form a statistically complete sample and as such may be subject to
considerable bias, e.g., we could be sampling the most luminous systems as the luminosity
of the accretion flow will have a spin dependence with larger spins facilitating more
luminous inner accretion flows.

Finally, we note that quasars are known to be a major contributor to the CXB, as such
knowledge of the precise form of their SED is important \citep{ueda14}. The reflection
contribution to the X-ray spectrum of \qso~is uncertain primarily due to our inability to
accurately constrain the iron abundance in the current observation, i.e., $\rm 0.5
\lesssim R_{frac} \lesssim 2.5$ (see Table \ref{model_table}). Further observations will
be required in order to constrain the iron abundance and hence the reflection fraction of
this high redshift black hole.

\acknowledgements We extend our thanks to the anonymous referee. This research has made
use of \textit{Chandra} data obtained from \textit{HEASARC}. This research made extensive
use of NASA's Astrophysics Data System.


\end{document}